# Network science based quantification of resilience demonstrated on the Indian Railways Network


Udit Bhatia[1], Devashish Kumar[1], Evan Kodra[1], Auroop R. Ganguly[1*]

[1]Sustainability and Data Sciences Lab, Department of Civil & Environmental Engineering, Northeastern University, Boston, MA 02115, USA.

\* Corresponding Author
E-mail: a.ganguly@neu.edu (ARG)




# Abstract


The structure, interdependence, and fragility of systems ranging from power-grids and transportation to ecology, climate, biology and even human communities and the Internet have been examined through network science. While response to perturbations has been quantified, recovery strategies for perturbed networks have usually been either discussed conceptually or through anecdotal case studies. Here we develop a network science based quantitative framework for measuring, comparing and interpreting hazard responses as well as recovery strategies. The framework, motivated by the recently proposed temporal resilience paradigm, is demonstrated with the Indian Railways Network. Simulations inspired by the 2004 Indian Ocean Tsunami and the 2012 North Indian blackout as well as a cyber-physical attack scenario illustrate hazard responses and effectiveness of proposed recovery strategies. Multiple metrics are used to generate various recovery strategies, which are simply sequences in which system components should be recovered after a disruption. Quantitative evaluation of these strategies suggests that faster and more efficient recovery is possible through network centrality measures. Optimal recovery strategies may be different per hazard, per community within a network, and for different measures of partial recovery. In addition, topological characterization provides a means for interpreting the comparative performance of proposed recovery strategies. The methods can be directly extended to other Large-Scale Critical Lifeline Infrastructure Networks including transportation, water, energy and communications systems that are threatened by natural or human-induced hazards, including cascading failures. Furthermore, the quantitative framework developed here can generalize across natural, engineered and human systems, offering an actionable and generalizable approach for emergency management in particular as well as for network resilience in general.




# Introduction

Complex networks have yielded novel insights across systems in ecology [1,2], climate [3], biology [4], transportation [5], water, power [6], and communications including the Internet [7,8] and social media [9]. Network science methods have resulted in actionable information on network vulnerabilities and fragility, especially in the context of response to disruptive events. Yet, network science methods have not often been used for post-perturbation recovery. In addition, they have not been used for developing a unified resilience framework that brings together event response with post-event recovery. However, a generalizable framework could help stakeholders of many complex systems design for, protect against, and recover from major disruptions. In this study, we develop such a framework in the context of natural and human-induced hazards using the *Indian Railways Network* (IRN) as an illustrative case study.

Large-Scale Critical Lifeline Infrastructure Networks (LSCLINs) include water distribution pipelines, power grids, telecommunication lines, railways, roadways, seaports, airports and communication networks. These LSCLINs are subject to growing threats from natural and man-made calamities, such as climate extremes, terrorism and cybercrimes. Aging of infrastructures, connectivity of lifeline functions, competition for resources, urbanization and movement towards coasts have exacerbated existing vulnerabilities. Resilience of LSCLINs has been recognized as an urgent societal imperative given the impact of recent hazards [10,11].

Multiple definitions and frameworks for assessing resilience have been proposed in literature [12]. A recent correspondence piece in *Nature* [13] points out that characterizing and measuring resilience can be a challenge given that it takes on more than 70 definitions in literature. In this study, we adopt the definition of resilience to be: "the ability to prepare and plan for, absorb, recover from, and more successfully adapt to adverse events" proposed by the



*National Academy of Sciences* [14], a definition that been further conceptualized in recent literature [10]. We note that while many operational, organizational and human factors [15] contribute towards systems resilience, the scope of this study is restricted to the analysis of network structure with an underlying assumption that these factors remain stable before, during, and after a hazard.

Literature has produced conceptual frameworks [10,11] for understanding resilience of LSCLINs. In addition, limited approaches for quantifying different aspects of resilience have been developed. Prior literature has proposed frameworks for quantification of resilience [16,17] with a focus on comparison of component-based retrofitting and structural restoration. However, as pointed out in [10], component-based risk management strategies may have limited utility for complex infrastructure system when diverse hazards need to be considered. One recent study proposed a framework for analyzing resilience based on modeling the functional dependencies among multiple systems, including critical infrastructures [18]. Another used a technique for measuring a system's ability to recover from disruptions; the method was based on stochastically ranking the importance of the system's components [19]. Yet another discussed quantifying the resilience of networked systems using network flow dynamics, metricizing resilience by using a "figure-of-merit" [20]. The figure-of-merit was essentially casted as a stand-in for any given specific metric describes the ability of a system to function successfully. The authors themselves [20] noted that while this metric theoretically characterizes a system's resilience, calculating it might not be trivial in real life settings, as it depends on parameters that may not actually be easy to obtain. These types of proposed models attempted to incorporate various realistic metrics, such as resistance of components to disturbance, duration of disturbance, counts of failing components [18], network vulnerability and survivability [19], and ability of a system to function



successfully [20]. Yet, overall, these approaches involve a number of parameters that may be hard to estimate or obtain, limiting their general applicability.

Other researchers proposed a three-stage resilience framework to quantify the resilience of networked urban infrastructures [16]. This research highlighted that while recovery sequences play a crucial role in resilience improvement in the context of finite resources, deploying redundancy, hardening critical components and ensuring rapid recovery are all effective responses regardless of their ordering to build resilience. Although network science frameworks were used to model the infrastructures, a pre-selected network measure was employed to generate the recover strategy in a resource-limited scenario. Another recent study [21] proposed a methodology for joint restoration modeling of interdependent infrastructure systems exemplified through interdependent gas and power system at county level impacted by hurricane. Here, genetic algorithm-derived scores were used to generate recovery sequences. As noted by the researchers, the proposed framework requires multiple models to function, including a hazard generation model, component fragility models and system performance models. We note that in both framework [16,21], damage and recovery of components (or nodes) was informed by disaster-specific fragility curves, which would not be readily available for many LSCLINs that face a variety of potential threats (perhaps unlike smaller urban or county scale networks).

We argue that the theory and tools provided by complex networks science can be developed or extended to produce a generalizable quantitative framework for characterizing and evaluating LSCLIN resilience. Network science has been used to understand the global air traffic networks, which have been delineated into communities that conveyed information beyond geopolitics, and examined through weighted networks for improved understanding of topology [22].Transportation networks at city scales have been examined through quantitative assessment



of fragility [23] but only with qualitative recovery strategies [24]. Air transportation systems have been modeled as dynamical complex networks to study the performances of air traffic managements systems under increasing traffic loads [25]. Cascading failure of links in power networks has been analyzed using random geometric graphs and percolation-based analysis [26]. Network science methods [7] coupled with computer technology [8] have been used to examine the robustness of the worldwide Internet network. The fragility of interdependent networks has been modeled and examined through simulations [27], including in a study where cascading failures across lifelines in Italy [28] were used as motivating examples. Another study [29] developed a generic network science-based approach to identify a system's response to a wide range of hazards; this study quantifies resilience using synthetic graphs and Linux software networks. That study proposed an approach that allows evaluation of resilience across time. The results demonstrated that how parameterizations for features such as redundancy, node recovery time, and backup supply available could be tuned to obtain a desired resilience state. While it suggests applicability of complex network-based algorithms for quantifying resilience, the proposed framework was only validated using synthetic and simulated graphs. Moreover, obtaining the parameter information to validate the proposed strategies for LSCLINs is a non-trivial task. A study proposed a network science based method for analyzing the resilience of the Chinese railway network as well as a method for optimizing the design of transportation networks, but not the methods for recovering existing ones [30].

To date, no study we are aware of has put forth an approach for translating recently proposed conceptual resilience curves [10] to practical tools for LSCLINs based on real world data. The network science centric framework offered in the current study is a data-driven embodiment of those conceptual resilience curves [10]. We show how it can be used to measure



the response of LSCLINs to multiple hazards as well as to generate and compare the effectiveness of restoration strategies in a quantitative and generalizable manner.

The previous literature [17,31] has examined restoration methods in the context of specific hazard types and infrastructure systems [33], often with known fragility models [17] or resource constraints [16] or other component level information , and occasionally with a recovery sequence predetermined through the selection of specific network metrics [16]. This paper presents a new generic data-driven version of the formerly conceptual resilience curves [10], which, as previously discussed, is particularly applicable to multi-hazard resilience of LSCLINs where fragility or component level information may not be available.

We illustrate this framework using the Indian Railways Network (IRN), the largest railways in the world in terms of passenger-kilometers transported per year. The IRN is among the most important lifelines in India. It transports more than 8.4 billion passengers annually and has played an important role in the nation's economy as well as in relief and rescue operations after both man-made and natural hazards [32]. For example, after 2013's Cyclone Phailin, the IRN played an important role across the eastern coast of the nation [33]. Given the economic and societal importance of this network, maintaining its functionality is crucial.

# Materials and Methods

**Indian Railways Network data**

Here, we analyze origin-destination data of passenger-carrying trains on the IRN. The network is constructed using publicly available data, which was cleaned and appropriately formatted prior to analysis. Open-source express and local passenger trains data are available on the following websites:



http://www.indianrail.gov.in/mail_express_trn_list.html, http://www.indianrailways.in/train-list.

This data is compiled by *Ixigo,* an e-ticket booking company and is publicly available on:

http://www.ixigo.com/trains/trains. We model the IRN as an origin-destination network. We

considered stations with at least one originating or terminating train, comprising a total of 809

stations with 7066 trains as of October 30, 2014. This study only considers origin and destination

stations, since data from stations that are between origins and destinations are usually not freely

and widely available. Only 752 of the 809 stations are part of the giant component (the largest

connected group of stations). We focus our analyses on this subset.

**Characterizing the IRN topology**

Each station's degree, or connectivity, is measured by the number of connections it has with

other stations. Each station's strength is measured by its traffic volume in terms of the total

number of trains that originate or terminate at that station. Strength is defined this way with the

hypothesis that traffic volume may be a useful metric for understanding failure and/or for

prioritizing stations during recovery. Two stations *i* and *j* are considered to be connected with an

edge if there exists a train between the pair of stations such that a train originating at *i* terminates

at *j*. Elements of the adjacency matrix $\{a_{ij}\}$ are 1 if the train originating from station *i* terminates

at station *j* and 0 otherwise. The weight of an edge is calculated as the number of trains running

between a pair of stations in either direction. Thus, any element of the weighted adjacency

matrix $\{w_{ij}\}$ is the number of trains originating from station *i* and terminating at station *j*. The

connections were almost all bidirectional; specifically, the numbers of trains connecting pairs of

stations in one direction were different from the other directions in less than 250 out of the

654,481 possible cases (less than 0.04% of cases). The traffic flow matrix could therefore be

made symmetric without much distortion of the network by selecting the larger non-zero value

per station pair. Hence, the IRN is analyzed as an undirected weighted network. To understand



the structure of IRN, we calculate the degree and strength distribution of the stations. The cumulative degree distribution $P(k > K)$ gives the probability that a station has *more than K* connections to other stations and is defined as:

$$P(k > K) = 1 - \sum_{k=k_{min}}^{K} p(k) \qquad (1)$$

where $p(k)$ is number of stations having degree $k$ divided by total number of stations and $k_{min}$ is the minimum degree found over all nodes in the network. Similarly, the cumulative distribution of strength $P(S > s)$ gives the probability that station has *more than S* originating (or terminating) trains, i.e., traffic volume. The cumulative degree and strength distributions follow truncated power law models. Most stations have a small number of connections, with the exception of several hubs that are generally related to major metropolitan areas and are geographically isolated from each other.

We also use the modularity-based [34] Louvain community detection algorithm to characterize the topology of the IRN. The weighted adjacency matrix defined earlier is used as the input for the community detection. Throughout the manuscript, we use the terms "community" and "module" interchangeably.

The topology of the network provides interpretation for the IRN's robustness to and recovery from different types and geographical origins of hazards, a topic that is discussed more in Results.

**Network robustness and recovery**

Quantifying resilience entails measuring the failure and recovery processes, which effectively requires the ability to measure the critical functionality of the IRN at any given state. Critical functionality is metricized as follows. We utilize the *giant component*, a common network science metric that measures the largest connected set of nodes in the network (i.e., in



the IRN defined here, within the giant component, one could travel from any station *i* to any other station *j* by at least one path) [35]. Total Functionality (TF) is the number of stations in the giant component when the network is completely functional; thus, in our case, TF=752. Fragmented Functionality (FF) is the number of stations in the giant component at any given step wherein one or more stations are incapacitated by disruptions (either during failure or recovery). We define State of Critical Functionality (SCF) =FF / TF; thus effectively, SCF is a measurement of critical functionality at any step normalized between 0 and 1. The SCF metric developed here, based on percolation theory [7], is similar to those developed in previous work on social networks [37].

Methods based on percolation theory have been widely used to understand the fragility of isolated systems such as the Internet [7], as well as interdependencies among the network multiplex[36,38]. However, the recovery strategies developed here differ from a straightforward application of percolation theory in that the sequence of node recovery does not necessarily follow the sequence in which the nodes were damaged during network collapse.

The network recovery sequences may be random (which can be particularly useful as a baseline for testing the effectiveness of non-random recovery sequences), based on station attributes (such as degree or strength), or based on network attributes (such as centrality).

Recovery takes on the following process:
1. SCF (as defined above) is computed at the initial post-hazard state.
2. A prioritization sequence is identified. This prioritization sequence is the order in which nodes should regain their full functionality. For example, restoring the node A to full functionality requires restoring all edges connected to the node and partially activating the nodes which are one step from node A. Nodes that are



partially activated may not have full functionality, since for these nodes, only the edges that directly lead to fully functional node are recovered. This sequence can be generated randomly, through intuitive metrics, or through network science metrics (discussed soon).

3. Given a sequence, iteratively until SCF=1:

   a. The next station in the prioritization sequence is restored to full functionality, re-establishing the traffic flow between this station and the stations to which it is connected. This grows the giant component.

   b. FF is recalculated.

   c. SCF is recalculated.

We note that time reversal asymmetry has been observed in the recovery of financial systems as a result of external forces as shown in [39]. However, the time reversal asymmetry in SCF in our algorithm, while apparently similar, has a different explanation and happens to be a consequence of the proposed recovery process (Fig. 1).



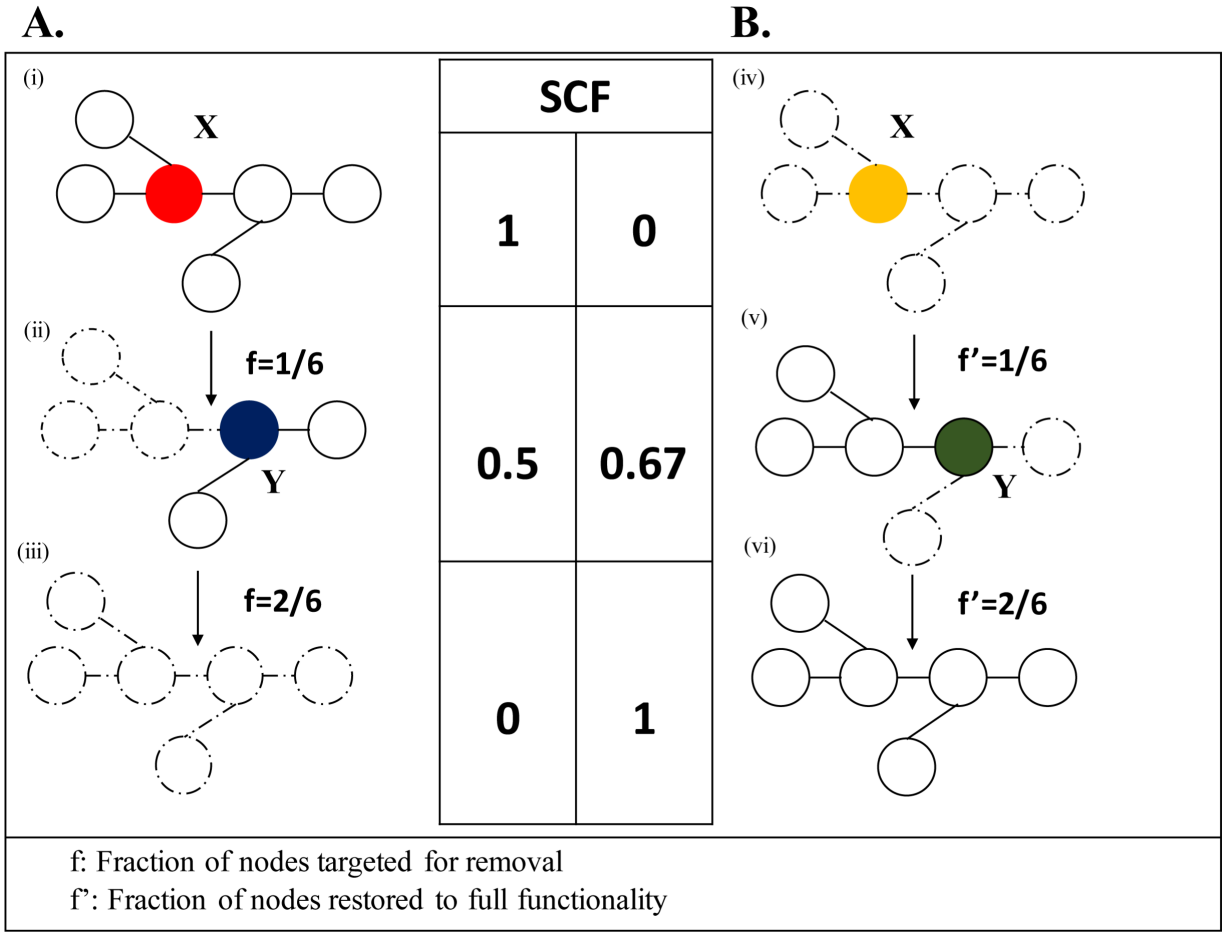

**Fig. 1 – Node removal and recovery process in the representative network with N=6**

Nodes X and Y are selected randomly for removal at time, T=1 and T=2, respectively. **A.** (i) The *SCF*=1 at step T=0 (pre-hazard). Node X (red) is selected for removal at step T=1. (ii) Removal of node X results in reduction of the size of the Giant Component (*GC*), which sets *SCF*=0.5. Dashed nodes (edges) means that nodes (edges) gets detached from the GC and hence incapacitated. Node Y (blue) is selected for removal at step T=2 (f=1/6, meaning one out of the six nodes is targeted for removal). (iii) The *GC* ceases to exist after the removal of node Y. **B.** To highlight the asymmetric nature of recovery process, nodes are restored to their full functionality in the same order these were removed (i.e. node X followed by node Y) from the network.



(iv) Node X (yellow) is selected for restoration to full functionality in the first step of the recovery process. (v) This results in the recovery of the node X to full functionality (f'=1/6, meaning one out of the six nodes is fully functional). As a result, three nodes directly connected to X gain at least one edge and the *GC* grows, making *SCF*=0.67. Then, node Y (green) is selected for recovery in step (vi). Recovery of node Y to its full functionality result in restoration of the SCF of the network to 1 as shown in (vi)

In practice, this is repeated with different sets of priority sequences in order to compare the effectiveness of different recovery strategies. Random sequences of node prioritization are used as a baseline to test the effectiveness of non-random strategies. We consider strategies based on descending order of degree and strength, which represent intuitive ("connectivity", and "traffic volume", respectively) recovery sequencing choices. Additionally, we consider several network centrality metrics: first, betweenness centrality of a particular node can be interpreted as the number of times a station acts as a bridge along the shortest path between two other stations. In the present context, it is the number of shortest paths connecting any two stations that involve a transfer at particular station. Betweenness centrality of each station, $B_k$, is normalized by dividing it by the network average betweenness centrality, $<B>$. Second, to understand the importance of stations due to its connections, we measure Eigenvector centrality. Here, a relative score is assigned to each station in the network, where a score is higher when its connections are themselves highly connected stations. Finally, closeness centrality of a node is calculated as the inverse of the average network distance of a given station to all other stations.

## Results

This section first presents results pertaining to IRN robustness and recovery, followed by an analysis and interpretation of IRN topology with implications for resilience.



**Robustness and recovery results**

Fig. 2A displays the IRN origin-destination network. As described in the Materials and Methods section, nodes are sized by traffic volume and colored according to community association. The 12-largest communities capture the vast majority (91.6%) of the railway stations. While the topology is discussed later in the Topology of the IRN and resilience implications subsection, Fig. 1A serves to provide context for the robustness and recovery results.

Fig. 2B (left panel) translates the formerly conceptual hazard response curves [10] into quantitative terms, with the IRN functionality degrading owing to either targeted station removal or random failures (RF) (e.g., potentially from random but typical service disruptions). The targeted station removals may be caused by targeted attacks (TA) that prioritize the stations to be taken down by connectivity or degree (Targeted Attack – Degree, TA-D), or by traffic volume or strength (Targeted Attack – Strength, TA-S). Network robustness computations applied to the IRN suggest that while RF would need to eliminate 95% of the stations for near complete loss of functionality; the corresponding numbers are 25% and 23% for TA-S and TA-D respectively.

In the right panel of Fig 2B, recovery strategies are compared for the case where the IRN starts at a *SCF*=0, i.e., completely unconnected and dysfunctional. Three types of recovery alternatives are evaluated. First, N=1000 random sequences serves as a baseline for comparison. The second set of strategies, perhaps the most immediately intuitive, is based on station attributes including connectivity and traffic volume. The third set of strategies is based on network centrality measures, specifically eigenvector (by average importance of connected stations), closeness (by average proximity, in a network connectivity but not necessarily a geographic sense, to other stations) and betweenness (by the average number of times any passenger traveling between origin-destination station pairs need to go through the station under consideration).



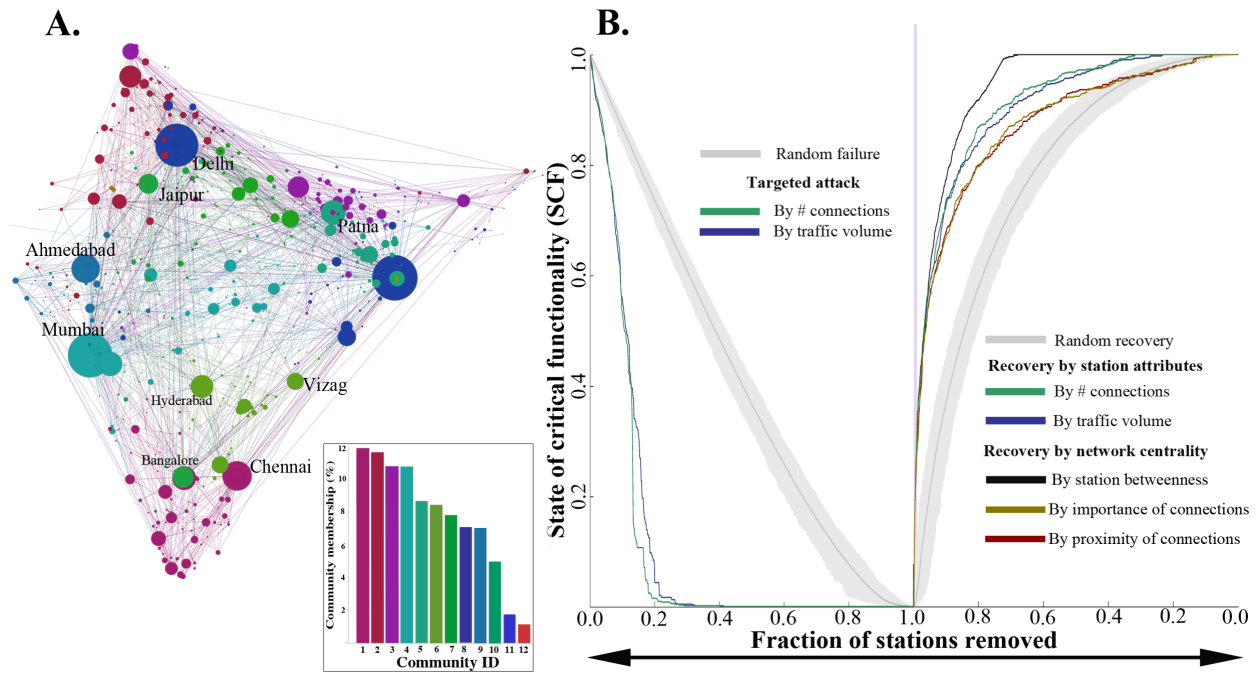

**Fig. 2. IRN and resilience curves under complete failure and recovery.**

**A.** The IRN is displayed. The 12-largest communities, each of which map to a color, capture the vast majority (91.6%) of the stations. Stations are sized by traffic volume. **B**. **(Left)** We quantify the robustness of IRN as it responds to random versus intentional attacks, where intentional attacks are motivated by either railway station connectivity (degree) or traffic volume (strength). For intentional attacks, approximately 20% of stations must be disrupted for the full IRN to lose all critical functionality, as measured with SCF. **(Right)** Using the same metric SCF, recovery strategies that propose alternative prioritizations for recovery of stations are compared, using an N=1000 ensemble of randomly generated sequences as a baseline. Number of connections (degree) and traffic volume (strength) are used as intuitive measures for generating recovery sequences, and the results are plotted. In addition, betweenness, Eigenvector, and closeness centrality are used, and the results are plotted.



In this case, IRN recovery is most efficient at most stages of partial or full recovery when betweenness centrality is chosen as for generating a recovery sequence. The efficiency of each recovery sequence is measured by computing its corresponding Impact Area [16,21] (IA). In the present case, IA is defined as the area between recovery curve and Y-axis representing SCF. Hence, a smaller IA signals a more efficient recovery strategy. On average, random recovery sequences have an IA (averaged over the N=1000 random sequences) that is at least 250% larger than the betweenness centrality-based sequence. The betweenness centrality-based sequence also has a 67% smaller IA than the sequence based on connectivity (degree). The interpretation for the performance of betweenness centrality in this case is discussed in the Topology of the IRN and resilience implications subsection.

The suitability of the framework for not only the full IRN but also its relatively independent communities (see Topology of the IRN and resilience implications subsection for discussion on communities) is studied. By proxy, the insights may generalize to LSCLINs in general and other systems, such as in food webs and ecological networks [40–43]. The two largest communities (as shown in Fig. 2A) were considered separately and analyzed in Fig. 3. While for full recovery betweenness centrality is the most efficient strategy, at some stages of partial recovery, the most efficient metric is less clear. For the community spanning South India, closeness centrality generates a particularly suboptimal sequence. This is likely a consequence of the fact that there are many geospatially proximal stations in this region that are not connected to many others. In both communities analyzed (South and North India), betweenness centrality ultimately emerges as the best metric for prioritizing stations for recovering the full IRN or the vast majority of its functionality. The results suggest that the framework generalizes but that the



choice of the most appropriate recovery strategy may depend on the network, the community, or the desired state of recovery (i.e., level of desired SCF).

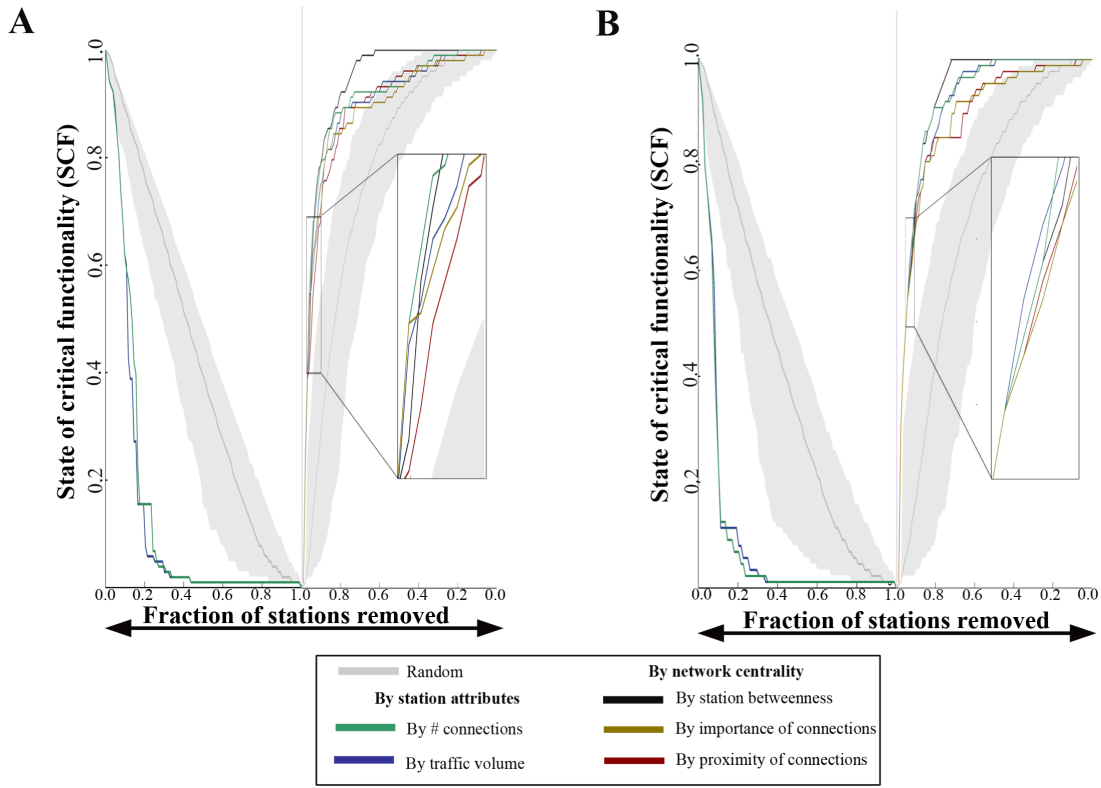

**Fig. 3 – Resilience curves for the IRN's two largest communities.**

The same as Fig. 2B is demonstrated for the two largest communities shown in Fig. 2A. **A.** The same as Fig. 2B is displayed but for the largest community (in South India, Community ID 1 from Fig 1B). **B.** The same as Fig. 2B. is shown but for the second largest community (in North India, Community ID 2 from Fig. 2A) with the inset showing that, at different levels of partial recovery (e.g., SCF ~= 0.4), it is not always clear which metric is most effective for prioritizing stations.

Although Figs. 1-2 readily provide a translation of previously conceptual temporal resilience curves [10] to those based on data, it may not be realistic for the network recovery process to begin from a state of complete disrepair (i.e., at SCF=0). This motivates testing the



utility of the framework on exemplary set of realistic hazards that only partially incapacitate the IRN.

We examine the recovery portion of the framework subject to three specific hazards. Fig. 4 illustrates the geographic characteristics of the three hazards as well as their impact on community structure; interpretation of the community structure is discussed more in the Topology of the IRN and resilience implications subsection. First, a simulation inspired by the 2004 Indian Ocean tsunami [44] removes 9% of stations on the southeastern Indian coast. Second, we simulate a scenario based on a cascade from the power grid, similar to the fallout from the historically massive 2012 blackout [45]. Finally, we simulate a cyber or cyber-physical attack scenario, where the stations are perhaps maliciously targeted based on traffic volume, and the network structure is fractured significantly. Consequences of physical- and cyber-attacks are based on hypothetical scenarios, although motivated from real-world events such as the terror attack of 26th November 2008 (the "26/11 Mumbai terror attack"), impacting the "Mumbai" station [46]. Cyber-physical terror attacks may be region-specific, but the possibility of coordinated attacks exists, especially for cyber. The post hazard SCF values are 0.903, 0.852, and 0.890 (679, 641, and 669 nodes remaining in the giant component) respectively for the three hazards, and recovery begins from these starting points. S1 Table shows which stations were removed for each of the three hazards.



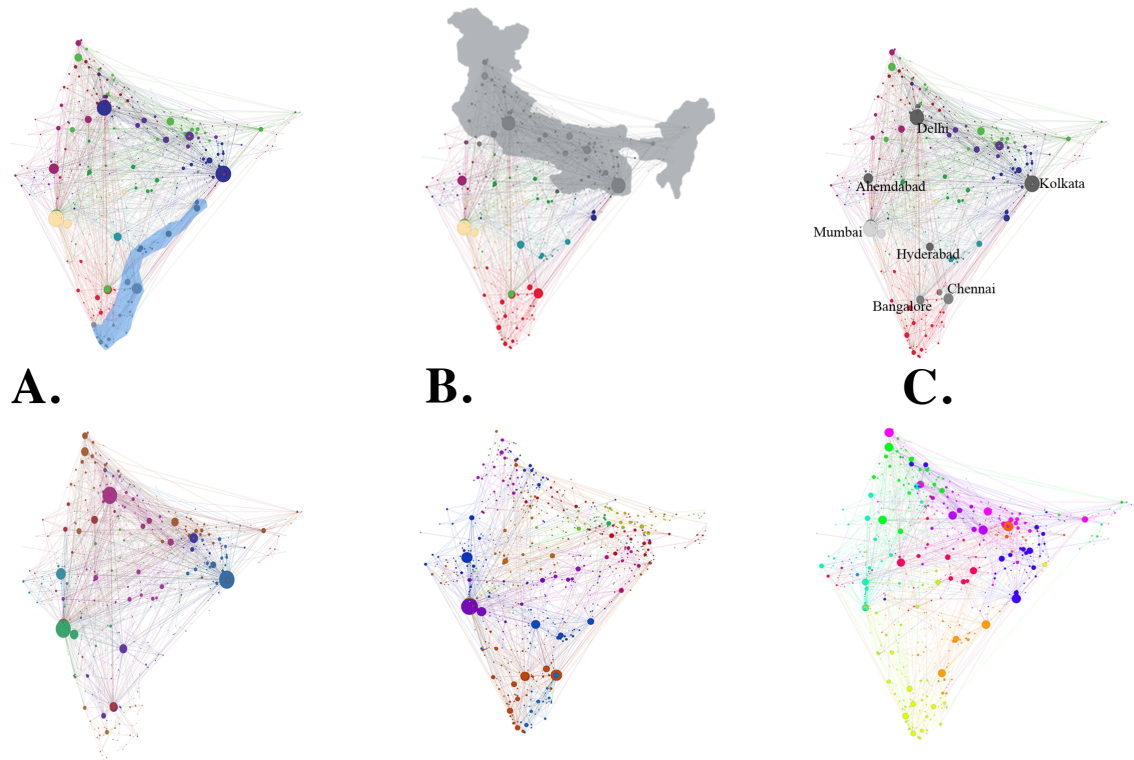

**Fig 4. Simulating the impact of realistic natural and man-made hazards.**

The top row schematically illustrates portions of the IRN initially impacted by realistic natural and cyber or cyber-physical threat scenarios, all with the same initial network topology as shown in Fig. 2A, where 752 stations reside in the largest giant component (SCF=1). The bottom row displays the community structure post hazard in each case. **A.** The impact of a disaster with properties similar to that of the December 2004 Indian Ocean tsunami is displayed. As suggested by the insight that communities are relatively independent as obtained from Fig. 2, the regional nature of the hazard (shaded blue, top) significantly impacts the Southeastern coast, removing 28 stations. The number of communities increases from 49 to 75. Yet, the structure of the remainder of the network remains relatively intact (SCF=0.903, where 679 stations remain in the giant component, see Methods and Data). **B**. For a simulated cyber or cyber-physical attack scenario, where 19 stations are perhaps maliciously targeted based on traffic volume (nodes shaded grey,



top) and removed, the network structure is fractured significantly (SCF=0.890, where 669 stations remain in the giant component). The number of communities increases from 49 to 96. **C**. A scenario based on a cascade from the power grid, similar to the 2012 blackout (shaded grey, top) is also simulated. The impact is significant, removing 39 stations, but the degradation of the IRN remains regionally contained (SCF=0.852, where 641 stations remain in the giant component). The number of communities increases from 49 to 102. Note that differences that appear relatively in the SCF can have significant practical implications with a large network. In the case of the IRN given TF=752, an SCF dropping by about 0.01 means 10 less stations are part of the giant component.

Fig. 5 A-C show recovery curves like those in Fig. 2B and Fig. 3 but starting from these states of functionality. Again, an ensemble of N=1000 members of random sequences are used in each case as a baseline. All metrics lead to recovery sequences that are almost always more effective than the random sequences. Fig. 5 D-F captures the percentage of random ensemble members that are exceeded by each strategy at each recovery step in terms of SCF. Here, it is less clear which metrics lead to the most effective recovery sequences in general. In all three cases, a choice of betweenness centrality leads to total functionality (SCF=1) earliest, although at earlier stages other metrics are at times preferable.



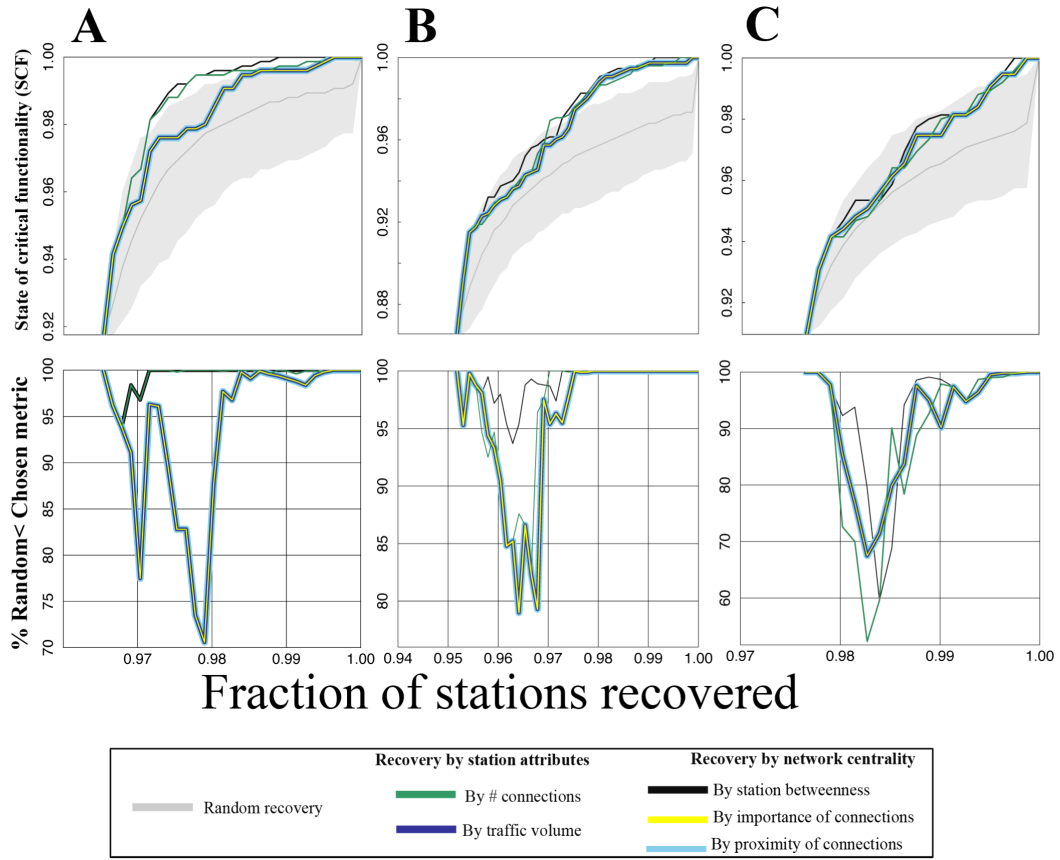

**Fig 5. Recovery from simulated natural and man-made hazards.**

**A.** Recovery curves after the simulated tsunami are displayed. As a baseline, the gray shaded interval represents the 99% bounds of N=1000 randomly generated recovery sequences. At each step, the 99% bounds are the 5th and 995th largest SCF values from the N=1000 member ensemble. The SCF begins at 0.903. **B.** The same as A is displayed but for the simulated cyber-physical attack. Here, the SCF begins at 0.852. **C**. The same as A is displayed but for the simulated power grid failure cascade. Here the SCF begins at 0.890. **D.** For the tsunami, at each step of the recovery curve, the percentage of ensemble members that a given metric is larger than in terms of SCF is plotted. This is repeated for each metric (connectivity, traffic volume, betweenness centrality, Eigenvector centrality, and closeness centrality). **E.** The same as D but



for the cyber-physical attack recovery curve. **F.** The same as D but for the power grid failure cascade. In D-F, in some cases, lines overlap each other; when this is the case, one line is thicker than the other to enable visibility of both.

**Topology of the IRN and resilience implications**

Analysis of the topological characteristics of the IRN lends a degree of interpretability to both the fragility and recovery components of the resilience results captured in Fig. 2, Fig 3, and Fig. 5.

First, Fig. 6A displays a cumulative probability distribution of stations' degree and strength (connectivity and traffic volume). They both follow truncated power distributions (see Materials and Methods). Several stations are hubs, but most have a small number of connections. The hubs tend to be *geographically* distant from each other. Table 1 provides additional details for the hubs shown here, delineating those that are labeled with the same city for brevity. Furthermore, Fig. 6B shows a correlation profile of the connectivity of the stations, which shows that stations with high connectivity do not tend to be connected to others with high connectivity. Hence, hubs tend to be not only distant from each other geographically but also in terms of *network* distance.



**Table 1– Network summary statistics on 25 most connected (highest degree) stations**

| Stations (*Metro regions**) | Degree | Strength | Closeness Centrality | Betweenness Centrality | Eigenvector Centrality |
|---|---|---|---|---|---|
| Howrah (*Kolkata*) | 77 | 133 | 0.41 | 0.11 | 1.00 |
| New Delhi (*Delhi*) | 62 | 126 | 0.39 | 0.06 | 0.84 |
| Delhi (*Delhi*) | 59 | 109 | 0.36 | 0.06 | 0.54 |
| LokManyaTilak(*Mumbai*) | 56 | 93 | 0.39 | 0.07 | 0.75 |
| Chennai | 49 | 86 | 0.39 | 0.06 | 0.71 |
| Ahmedabad | 48 | 83 | 0.38 | 0.06 | 0.76 |
| Mumbai (*Mumbai*) | 46 | 130 | 0.37 | 0.04 | 0.51 |
| Yesvantpur (*Bengaluru*) | 44 | 62 | 0.38 | 0.04 | 0.65 |
| Hazrat Nizammudin (*Delhi*) | 42 | 65 | 0.37 | 0.03 | 0.63 |
| Pune | 41 | 70 | 0.38 | 0.03 | 0.62 |
| Bangalore (*Bengaluru*) | 39 | 69 | 0.36 | 0.04 | 0.44 |
| Patna | 39 | 72 | 0.37 | 0.04 | 0.52 |
| Secunderabad | 39 | 66 | 0.37 | 0.05 | 0.52 |
| Jaipur | 37 | 57 | 0.36 | 0.03 | 0.53 |
| Amritsar | 36 | 64 | 0.36 | 0.03 | 0.50 |
| Puri | 36 | 53 | 0.37 | 0.03 | 0.58 |
| Jammu Tawi | 35 | 47 | 0.37 | 0.02 | 0.60 |
| Varanasi | 35 | 51 | 0.37 | 0.03 | 0.50 |
| Ajmer | 34 | 42 | 0.37 | 0.02 | 0.58 |
| Bandra | 34 | 50 | 0.35 | 0.01 | 0.50 |
| Gorakhpur | 33 | 62 | 0.36 | 0.03 | 0.50 |
| Visakhapatnam | 32 | 49 | 0.35 | 0.02 | 0.35 |
| Tirupati | 31 | 49 | 0.34 | 0.02 | 0.36 |
| Indore | 30 | 38 | 0.36 | 0.02 | 0.48 |
| Kolkata (*Kolkata*) | 29 | 36 | 0.35 | 0.02 | 0.43 |

*Metro regions are indicated within parentheses to delineate hubs shown more than once in Fig. 5A

Referring back to Fig. 2A, communities tend to be associated with one or a few large stations and are typically guided by geographical distance as well as by cultural or linguistic closeness. Thus, while most communities tend to be regionally grouped, notable exceptions to this rule exist, reflecting high connections among larger but geographically distant stations, notably near Delhi and Kolkata. Since the communities are detected based on systematic analysis of traffic flow properties, they serve as a more appropriate characterization of modular structure



than more conventional – but from the network perspective, arbitrary – divisions, such as railways jurisdictions or political boundaries.

Fig. 6 A-B, in conjunction with the dominance of the geographic grouping structure revealed by the community detection algorithm shown in Fig. 2A, imply that the IRN as a whole tends to function like a group of loosely connected, relatively independent modules.

These topological features provide intuition into recovery results. For the results shown in Fig. 2B and Fig. 3A-B, namely those where the IRN is recovered from complete dysfunction, betweenness centrality outperforms other metrics at most stages of partial recovery. That the IRN generally tends to function as a group of relatively independent modules lends interpretation here: stations with higher values of betweenness centrality act as key bridges between those modules. Hence, recovering those key bridge stations earlier in the sequence help rapidly grow the giant component by connecting groups that will be large but otherwise unconnected. Finally, Fig. 6C shows rank correlation for each centrality metric with degree and strength. Correlations are positive and increasing, but rank orders are substantially different. This means that sequences generated from these metrics differ substantially as well, providing insight into the observed wide gaps often seen betweenness centrality-generated recovery curves and others, including degree and strength.

The best performing metric is less clear for recovery from the three simulated hazards. While all proposed metrics almost always outperform the randomly generated recovery sequences, the IRN is not nearly as fragmented when recovery begins. Modules are not as disconnected, and hence betweenness centrality is no longer clearly preferable to other metrics.



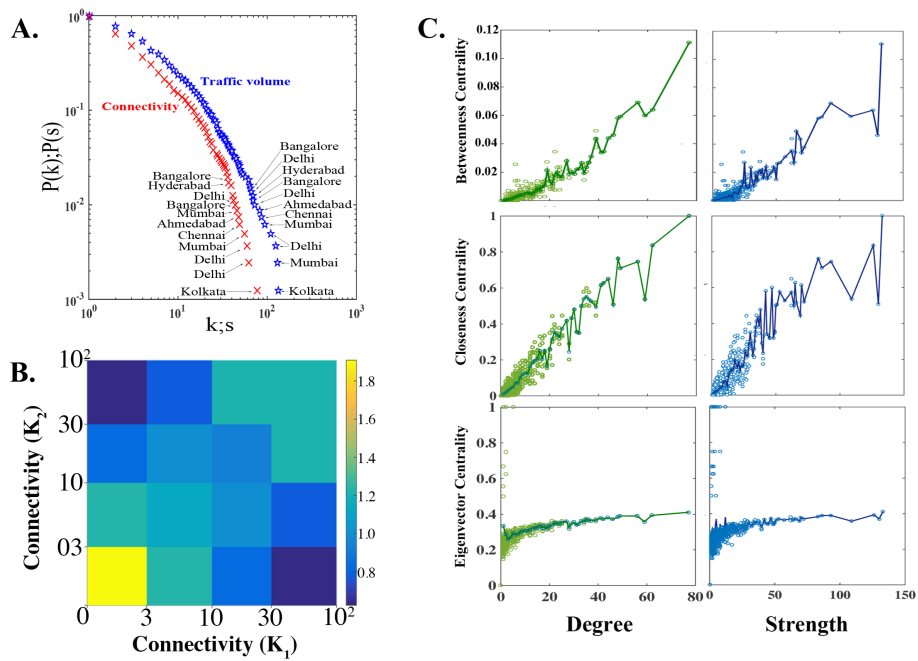

**Fig. 6 – Topology of the IRN**

**A.** A cumulative probability distribution of node degree and strength (as measured by traffic volume) of stations in IRN, on a log-log scale, profile the distributional properties of the stations. The distributions follow truncated power law models, wherein most stations have a small number of connections, with the exception of a few hubs. Hubs are generally geospatially isolated. "k" stands for degree, and "s" stands for strength. Several cities are labeled multiple times as they contain more than one hub. For example, Delhi actually has multiple hub stations, specifically stations named "Hazrat Nizammudin", "New Delhi" and "Delhi"; Delhi was used for brevity to represent all three. Table 1 details and delineates all stations that have been named identically in this panel. **B.** A correlation profile of station connectivity of IRN shows the average degree of stations' nearest network neighbors. $K_1$ and $K_2$ serve to index the degree of any given station. Correlations in connectivity are shown as systematic deviations of the ratio $P(K_1,K_2)/P_r(K_1,K_2)$. $P(K_1,K_2)$ is the likelihood that two stations with connectivity $K_1$ and $K_2$ are connected to each



other by the direct link. $P_r(K_1,K_2)$ is the same value in averaged over a randomized ensemble of 1000 members. Yellow colors in the lower left indicate the tendency of stations with less connectivity to connect to other stations with comparable connectivity, while blue/green colors indicate small likelihood of hubs connecting with one another indicating the IRN's disassortative nature. This further captures the tendency of the IRN to behave like a collection of relatively independent modules. **C.** Degree and strength are plotted against betweenness, closeness and Eigenvector centrality. Lines indicate the average for a centrality measure conditional on a particular degree or level strength, respectively, serving to highlight the variability in centrality metrics even for identical levels of connectivity or traffic volume.

## Discussion

This study presents a complex networks-based unified framework that goes beyond fragility characterization and conceptual resilience curves to offer data-driven, quantitative insights for decision-making before, during and after hazards to enable preparedness, relief and recovery.

While the fragility characterization may be considered an adaptation of existing network science methods, recovery curves are new and can generalize to other applications. The framework developed here allows for generation and performance comparison of multiple station recovery sequences, allowing for the possibility that different networks should be recovered according to sequences generated from different metrics. The recovery differs from a straightforward application of percolation theory in that the sequence of node recovery does not necessarily follow the sequence in which they were damaged during network collapse. The recovery of a node accompanies the recovery of the links and hence traffic flow to directly connected nodes.



This study also characterizes the topology of the IRN, linking insights from the analysis to the performance of best performing recovery strategies. In this case, betweenness centrality tends to generate the best performing recovery sequences due to the inherent modularity of the system. Recovering stations that serve as key bridges between these modules tend to bring functionality online most effectively. It is not clear whether betweenness centrality-based sequences would perform best for all LSCLINs; however, topological analyses similar to the one conducted in this study may help provide interpretability for best performing metrics for other LSCLINs as well.

Anticipatory analysis may help stakeholders design systematic recovery strategies for LSCLINs including transportation, water and wastewater, power and fuels, and communications systems. In addition to engineered systems such as LSCLINs, the approach can be generalized to natural systems such as ecological networks [43] subjected to perturbations [40,41].

Future extensions to the network resilience framework may need to consider the formulation of an efficient algorithm to approximate the optimal recovery strategy in light of recent literature on influence maximization [47] based on percolation theory [38,48,49] . Inclusion of link damage or removal information or metadata, resource mobilization strategies, structural redundancies [50], organizational, social and human factors as well as the consideration of dynamic network flow properties, including time varying network attributes [3] and real-time data ingestion could be valuable avenues to explore in future extensions of this study.



# Acknowledgments

Funding was provided by Northeastern University and the US National Science Foundation's (NSF) Expeditions in Computing Award #1029711, and NSF BIGDATA award # 1447587. The open source software *Gephi* was used for network visualizations. The authors thank the following Northeastern University colleagues for helpful discussions: R. Sinatra and A.L. Barabási on network science, J.F. Hajjar and S. Flynn on infrastructural engineering and policy respectively, as well as D. Wang and T. J.Vandal for comments on the manuscript.

# Supporting Information

**S1 Table. List of the stations removed for each of the three hazards**



**S1 Table: List of the stations removed for each of the three hazards**

| Tsunami | Power Failure | Cyber-Physical |
|---|---|---|
| Bhubaneswar | Ajmer | Ahmedabad |
| Chennai Central | Ambala Cant | Bangalore City |
| Chennai Egmore | Amritsar | Chennai Central |
| Cuttack | Anand Vihar | Chennai Egmore |
| Ernakulam | Asansol | Delhi |
| Gudivada | Bareilly | H Nizamuddin |
| Guntur | Bareilly City | Howrah Jn |
| Kanyakumari | Bikaner | Hyderabad Decan |
| Kochuveli | Danapur | Jaipur |
| Kollam Jn | Darbhanga | Kacheguda |
| Machilipatnam | Delhi | Kolkata |
| Madurai | Delhi S Rohilla | Lokmanyatilak |
| Mayiladuthurai | Dibrugarh | Mumbai |
| Nagercoil | Dibrugarh Town | New Delhi |
| Narasapur | Firozpur Cant | Pune |
| Puducherry | Ghaziabad | Sealdah |
| Puri | Gorakhpur | Secunderabad |
| Rameswaram | Guwahati | Shalimar |
| Sengottai | H Nizamuddin | Yesvantpur |
| Tiruchendur | Howrah | |
| Tiruchirapalli | Jaipur | |
| Tirunelveli | Jammu Tawi | |
| Tirupati | Kamakhya | |
| Trivandrum | Kanpur Anwrganj | |
| Tuticorin | Kanpur Central | |
| Vijayawada | Kolkata | |
| Villupuram | Lal Kuan | |
| Visakhapatnam | Lucknow | |
| | Lucknow | |
| | Muzaffarpur | |
| | New Delhi | |
| | New Tinsukia | |
| | Palwal | |
| | Patna | |
| | Pratapnagar | |
| | Puri | |
| | Sealdah | |
| | Tinsukia | |
| | Varanasi | |